\documentclass[twocolumn,showpacs,aps,prl,superscriptaddress]{revtex4}
\usepackage{graphicx}
\usepackage{dcolumn}
\usepackage{amsmath}
\usepackage{color}

\usepackage{amssymb}
\input pubboard_babarsym_mod

\newcommand {\GCXDzlnu           } {\ensuremath{  \Dz\electron\Anu              }\xspace}

\newcommand {\BzbtoDstarpennueb}  {\ensuremath{\Bzb\to\Dstarp\en\nueb}\xspace}
\newcommand {\BmtoDstarzennueb}   {\ensuremath{\Bm\to\Dstarz\en\nueb}\xspace}

\newcommand {\dm}         {\ensuremath{\Delta{}m}\xspace}
\newcommand {\cosby}      {\ensuremath{\cos\theta^{*}_\mathrm{BY}}\xspace}
\newcommand {\wtilde}     {\ensuremath{\tilde{w}}\xspace}

\newcommand {\rhosqraone}     {\ensuremath{\rho^{2}_{A_1}}\xspace}

\newcommand{\FoFaF}           {\ensuremath{         F }}

\newcommand{\dgdw}[1]        {\ensuremath{  \frac{\mathrm{d}\Gamma\left({#1}\right)}{\mathrm{d}w}  }}

\newcommand {\BrFr}[1] {\ensuremath{\mathcal{B}\left(#1\right)}\xspace}

\definecolor{Sig}{rgb}{1,1,1}
\definecolor{Dssdmp}{rgb}{0.0,1.0,1.0}
\definecolor{Dssdmf}{rgb}{0.51953125, 0.7578125, 0.63671875}
\definecolor{Cor}{rgb}{1,1,0}
\definecolor{Uncor}{rgb}{0,0,1}
\definecolor{Semisig}{rgb}{0.74609375, 0.5078125, 0.46875}
\definecolor{Dlnu}{rgb}{1.0, 0, 1.0}
\definecolor{Combds}{rgb}{0, 1, 0}
\definecolor{Ccbar}{rgb}{0.45703125, 0.5390625, 0.56640625}

\newcommand {\rd}{{\rm d}}

\newcommand {\ri}{{\rm i}}

\newcommand {\rB}{{\rm B}}

\newcommand {\rD}{{\rm D}}

\newcommand {\rK}{{\rm K}}

\newcommand {\Begeq}{\begin{equation}}
\newcommand {\Endeq}{\end{equation}}
\newcommand {\bEa}{\begin{eqnarray}}
\newcommand {\eEa}{\end{eqnarray}}

\newcommand{\grad}{\ensuremath{^\circ}}

\newcommand{\BABARPubYear}    {07}
\newcommand{\BABARPubNumber}  {002}

\newcommand{\SLACPubNumber}   {12667}

\def\figurebox#1#2#3{%
    \def\arg{#3}%
    \ifx\arg\empty
    {\hfill\vbox{\hsize#2\hrule\hbox to #2{\vrule\hfill\vbox to #1{\hsize#2\vfill}\vrule}\hrule}\hfill}%
    \else
    {\hfill\epsfbox{#3}\hfill}%
    \fi}

\begin{document}
%\linenumbers
%\linenumbers % Zeilennummern einschalten
%\modulolinenumbers[2] % jede 2.Zeile

\preprint{\babar-CONF-\BABARPubYear/\BABARPubNumber} 
\preprint{SLAC-PUB-\SLACPubNumber} 

\begin{flushleft}
\babar-CONF-\BABARPubYear/\BABARPubNumber\\
SLAC-PUB-\SLACPubNumber\\
%hep-ex/\LANLNumber\\[10mm]
%\mbox{\normalsize {\babar\ }Analysis Document \#\BADNumber, Version \BADNumberVersion}
\end{flushleft}

\title{
{\large \bf
Measurement of the Decay \BmtoDstarzennueb}
}

% Dummy author list; contact PubBoard Chair for current author list
%% author list as of 06-Jun-2007 (572 authors)
%
\author{B.~Aubert}
\author{M.~Bona}
\author{D.~Boutigny}
\author{Y.~Karyotakis}
\author{J.~P.~Lees}
\author{V.~Poireau}
\author{X.~Prudent}
\author{V.~Tisserand}
\author{A.~Zghiche}
\affiliation{Laboratoire de Physique des Particules, IN2P3/CNRS et Universit\'e de Savoie, F-74941 Annecy-Le-Vieux, France }
\author{J.~Garra~Tico}
\author{E.~Grauges}
\affiliation{Universitat de Barcelona, Facultat de Fisica, Departament ECM, E-08028 Barcelona, Spain }
% ...
\collaboration{The \babar\ Collaboration}
\noaffiliation

\date{\today}% It is always \today, today, but you may specify any date with \date.

\begin{abstract}
Using 226 million \BB events recorded on the
\FourS resonance with the \babar{} detector at the SLAC \epem
storage rings \pep2,
we reconstruct \BmtoDstarzennueb decays using the decay chain
$\Dstarz\to\Dz\piz$ and $\Dz\to{}\Km\pip$.
From the dependence of their differential rate on $w$, 
the product of the four-velocities of \Bm and \Dstarz,
and using the description of the
form factor $\FoFaF(w)$ by Caprini et al.,
we obtain the preliminary results 
$\rhosqraone        = 1.15\pm 0.06 \pm 0.08$, 
$\FoFaF(1)\cdot\Vcb =(36.3\pm 0.6  \pm 1.4)\cdot 10^{-3}$, and 
$\BrFr{\BmtoDstarzennueb} = (5.71\pm 0.08\pm 0.41)\%$. The first errors
are statistical and the second ones are systematic.
\\ \ 
\begin{center}
{\it Submitted to the 2007 Europhysics Conference on High Energy Physics,\\
Manchester, England.}
\end{center}
\end{abstract}

% PACS, the Physics and Astronomy Classification Scheme.

\maketitle

The exclusive \B-meson decay modes with the highest rates are the two
semileptonic modes \BzbtoDstarpennueb{} and \BmtoDstarzennueb{}.
Whereas the first has been measured by many experiments
\cite{Barberio:2007cr} to determine its rate $\Gamma$, its
differential rate $\rd\Gamma/\rd w$, and the CKM matrix element \Vcb,
the second has only been measured by two groups
\cite{Adam:2002uw,Albrecht:1991iz} with lower statistics. 
In the \Bz mode, the observed differential decay rate is
well described by heavy-quark effective QCD (HQET) using form factors
with the slope parameter $\rho^2$. 
However, the \Bz experiments do not agree well in their $\rho^2$
results.
Using the isospin symmetry
${\rm d}\Gamma (\BmtoDstarzennueb) =  {\rm d}\Gamma
(\BzbtoDstarpennueb)$, a precision measurement for the \Bm mode can
help to improve knowledge of $\rho^2$ and consequently of $\Gamma$
and  \Vcb.

The aim of our analysis \cite{JensThesis} is the determination of
the differential decay fraction $\rd{\cal B}/\rd w
(\BmtoDstarzennueb)$, where $\cal B$ is related to the decay rate
$\Gamma$ by the known lifetime $\tau(\Bm)$ and $w$ is the invariant
product of the four-velocities of \Bm and \Dstarz. The
neutrino in the \BmtoDstarzennueb decay is not reconstructed.
Therefore, the $w$ value of each reconstructed event cannot be obtained, only an approximation
\wtilde which will be defined below. Instead of unfolding, the parametrized
$\rd{\cal B}/\rd w$ expectation together with the $w$ resolution from
Monte Carlo simulation (MC) is fitted to the observed $\rd{\cal B}/\rd {\wtilde}$
distribution. 
Our fit, as in other recent $\Bz\to\Dstar\ell\nu$ analyses, uses the
parametrization of Caprini et al.
\cite{Caprini:1997mu}
and determines the two parameters $\FoFaF(1)\cdot\Vcb$ and $\rho^2$.
The third result, ${\cal B}
(\BmtoDstarzennueb)$,
is obtained by integrating $\rd{\cal B}/ w$. The parametrization is defined by
the following expressions:
\begin{widetext}
\begin{eqnarray*}
&&\dgdw{\BmtoDstarzennueb}
=\ 
 \frac{ G_F^2  } {48{\pi}^3}
 \left( m_{B} - m_{\Dstar} \right)^2 m_{\Dstar}^3 \sqrt{w^2-1} \left( w+1 \right)^2 \\
&&
\phantom{\dgdw{\BmtoDstarzennueb}
=\ }
\times \left[ 1 + \frac{4w}{w+1} \frac{m_{B}^2-2w m_{B} m_{\Dstar}+ m_{\Dstar}^2}{\left(m_{B}-m_{\Dstar}\right)^2} \right] \cdot \left .\FoFaF(w)\right .^2 {\Vcb }^2,\\
&&\left .\FoFaF(w)\right .^2 =
\left| h_{A_1}(w) \right|^2
\left[ 1 + \frac{4w}{w+1} \frac{m_{B}^2-2w m_{B} m_{\Dstar}+ m_{\Dstar}^2}{\left(m_{B}-m_{\Dstar}\right)^2} \right]^{-1}
\sum_{i=0,+,-}  \left|  \tilde{H}_i(w)\right|^2,\\
\end{eqnarray*}
\begin{equation*}
\left| \tilde{H}_0(w) \right|^2
=
\left[ 1+ \frac{w-1}{1-r} \left(1-R_2(w)\right) \right]^2,
\qquad
\left| \tilde{H}_{\pm} (w)\right|^2
=
\frac{1-2wr+r^2}{(1-r)^2}
\left[ 1 \mp \sqrt{\frac{w-1}{w+1}}R_1(w) \right]^2,
\end{equation*}
\begin{equation*}
\frac{h_{A_1}(w)}{h_{A_1}(1)}
=
1
-8\rho^2_{A_1}{}z
+\left(53\rho^2_{A_1}-15\right)z^2
-\left(231\rho^2_{A_1}-91\right)z^3,
\qquad
z  = \frac{\sqrt{w+1}-\sqrt{2}}{\sqrt{w+1}+\sqrt{2}},
\end{equation*}
\begin{equation*}
R_1(w)  =  R_1(1) - 0.12(w-1) + 0.05(w-1)^2,
\qquad 
R_2(w)  =  R_2(1) + 0.11(w-1) - 0.06(w-1)^2.
\end{equation*}
\end{widetext}
Note that $\rho^2=\rhosqraone$ in this notation. The parameters $R_1(1)$ and $R_2(1)$
are not determined in this analysis; we use the \babar{} results
from the $\Bzb\to\Dstar\ell\nu$ decay as input, see
\cite{Aubert:2006mb} and Table \ref{tbl:result:input_parameters}.

For our analysis, we use 205\invfb of \epem annihilation data recorded at 
$\sqrt{s}\approx{}m(\FourS)$ with the \babar{} detector
\cite{Aubert:2001tu}
at the SLAC storage rings PEP-II \cite{PepII}.
In addition to these on-peak data, we also use 16\invfb of off-peak data
collected below the \FourS resonance.
We select \BmtoDstarzennueb candidates \cite{Charge_Conj_Decays}
by pairing electrons with $p^*>1.2${}\gevc in the \epem rest frame
with \Dstarz candidates.
Since the precision of our results is not limited by statistics, we restrict
the analysis to the sequential decay modes $\Dstarz\to\Dz\piz$ and
$\Dz\to\Km\pip$, which have the smallest combinatorial background
and the best resolution in $\dm\equiv{}m(\Km\pip\piz)-m(\Km\pip)$.

Charged particles are selected if they have at least 10 hits
in the \babar{} drift chamber, transverse momentum $\pt>0.1${}\gevc, and a polar angle between
23.5\grad{} and 145.5\grad. Electrons (kaons) are selected with tight (loose) \babar{} particle
identification criteria. Neutral pions are reconstructed from two photons with energy
above 30\mev and a photon-compatible lateral shower shape in the \babar{} calorimeter. The
invariant mass must be $115 < m_{\gamma\gamma} < 150\mevcc$, and the photon pair
is then constrained to a common vertex and to
$m_{\gamma\gamma} = m(\pi^0)$. The decay candidates have to fulfill
the following further requirements:
The \Dstarz{}-\Dz mass difference must be
$135<\Delta m<153\mevcc$ and the \Dz{}-candidate mass
$1.8496<m(\rK^-\pi^+)<1.8796\gevcc$.
To reject non-B-decay candidates, the normalized Fox-Wolfram moment
$R_2$ \cite{Fox:1978vw} of the event
has to be smaller than 0.45. To reject candidates with a \Dstarz from one \B meson and
an electron from the other \B in the event, the angle between the \Dstarz and the \en has to be
larger than 90\grad.

Since there are many low-energy background photons, the selection criteria result
in many events with two or more
$\Dstarz\electron$ candidates, on average 1.75 per event. All $\Dstarz\electron$ candidates
in the same $\electron{}K\pi$ combination are
collected into one candidate group; on average there are 1.015 candidate
groups per event. Only one candidate
group per event is kept, in case of multiple groups the one with the smallest deviation
$|m(K\pi)-m(\Dz)|$. All
candidates in one group are kept in the analysis because the simulation of
low-energy photons is not perfect.
This procedure ensures that correctly reconstructed candidates are
selected with the same probability in data and MC.

The set of surviving candidates is binned in three dimensions according to their
values of \dm, \cosby, and \wtilde. The first two variables are used for the
separation between signal and background, the third is used for
the $w$ dependence of the signal. \dm is defined above, and
$\theta_{\rm BY}^*$ is the angle between the directions
of the \B meson and the $Y=\Dstarz+e$ system in the \epem rest frame
under the hypothesis that the \B decays
into only \Dstarz, \electron, and neutrino. It is defined by the four-vector relation
$$p_\nu^2=0=m_\rB^2+m_{\rm Y}^2-2(E_\rB^* E_{\rm Y}^*
                  -|{\vec p}_\rB^{\,*}||{\vec p}_{\rm Y}^{\,*}|\cos\theta_{\rm BY}^*)\ .$$
The value of
$$w=w(\beta^*)\equiv\frac{E_\rB^* E_{\rD^*}^*-
               |{\vec p}_\rB^{\,*}||{\vec p}_{\rD^*}^{\,*}|\cos\beta^*}{m_\rB m_{\rD^*}}$$
cannot be determined since the angle $\beta^*$ between the \B and the \Dstarz
in the \epem rest frame is
unknown. However, $\beta^*$ is bound between a minimum and a maximum value, and
$${\tilde w}=[w(\beta^*_{\rm min})+w(\beta^*_{\rm max})]/2$$
is a good estimator for $w$ in each event. $w$ and $\tilde w$ span a range from 1 to
1.5, the distribution of ${\wtilde}-w$ is
nearly Gaussian with an RMS of  0.026. We use 10 equidistant bins of \wtilde,
their width corresponds to about 2 RMS.

The fit for the two parameters $V=\FoFaF(1)\Vcb$ and $\rho^2$
is a binned maximum-likelihood fit in three dimensions.
The fit function is the sum
of the expected signal function $S=S(V,\rho^2)$ and the various
expected background functions. The signal
function in each $\tilde w$ bin is 
taken as the product of one-dimensional functions of \dm and \cosby. These
two functions are obtained from fits to the reweighted
signal MC distributions with
$V$-, $\rho^2$-, \linebreak[4]{$R_1(1)$-,} and $R_2(1)$-dependent weights on the generator level.
The signal fit function also
includes the normalization to the total number of $226\times{}10^6$ produced \BB pairs,
all decay fractions of sequential decays, the \Bm lifetime, all MC
reconstruction efficiencies, and efficiency corrections derived from control data
samples and their MC expectation. Efficiency corrections for track reconstruction
and charged particle identification follow those of other recent \babar{} analyses.
For the correction of the \piz reconstruction efficiency we use a control sample of $\tau$-lepton
decays as described below.
Small corrections are also applied for deviations of the shapes of the \dm distributions
in data and MC
because of track resolution differences, and for deviations in the shapes of
the \cosby distributions because of storage-ring energy calibration and resolution.

The background expectation functions are separately determined
for 23 classes of backgrounds.
This large number of background functions was necessary in order
to express each function $B_{\ri,{\tilde w}}$ as the product of
$B_{1,\ri,\wtilde}(\dm)$ and  $B_{2,\ri,\wtilde}(\cosby)$.
The one-dimensional fit functions $B_{j,\ri,\wtilde}$ are again
obtained from fits to MC distributions.
The fit to the data has 49 free parameters, in addition to $V$ and
$\rho^2$ there are 47 for adjustments of background normalizations and shapes, \dm
shapes, and \cosby shapes.

Before fitting the expectation function to the data, it is fitted to five different MC
subsamples whose size corresponds to the one of the data sample.
All five results for $V$ and $\rho^2$
agree with the MC input by better than one standard deviation. When applied to the data,
the fit result is $V=(36.32\pm 0.60)\cdot 10^{-3}$ and $\rho^2=1.146\pm 0.055$ with
a correlation coefficient $\varrho=+0.90$. Integrating
$\rd{\cal B}/\rd w$ over all $w$ leads to ${\cal B}=(5.71\pm 0.08)\%$. 
The total number of signal events is found to be $23\,499\pm329$.
Though the fit
is maximum-likelihood, a control value of $\chi^2$ can be calculated after the fit as
a goodness-of-fit measure. We find 4436.3 for 4095 degrees of freedom which is,
purely statistically, 3.8 $\sigma$ too high. Inspecting the distribution of per-bin
contributions to $\chi^2$ in all bins of \wtilde, \dm, and
\cosby, we find no concentrations of high values in any area.

\begin{figure}
\begin{center}
\begin{minipage}{0.45\textwidth}
\centering \includegraphics[width=1.0\textwidth,clip,bb=  10 0 516 349]{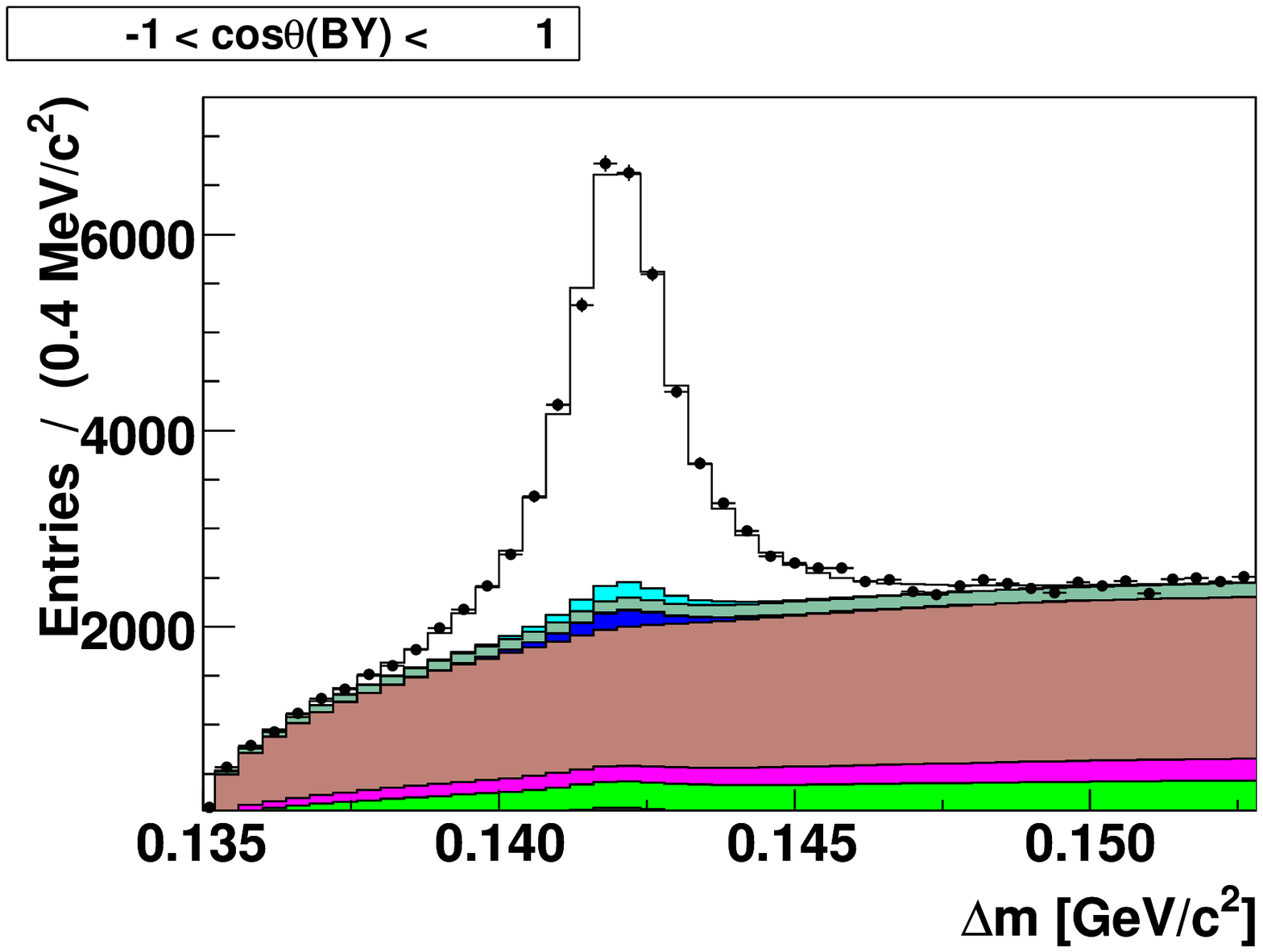}
\put(-20,145){(a)}
\put(-70,125){\lbabar\ }
\put(-70,115){{preliminary} }
\end{minipage}
\begin{minipage}{0.45\textwidth}
\centering \includegraphics[width=1.0\textwidth,clip,bb=  10 0 516 349]{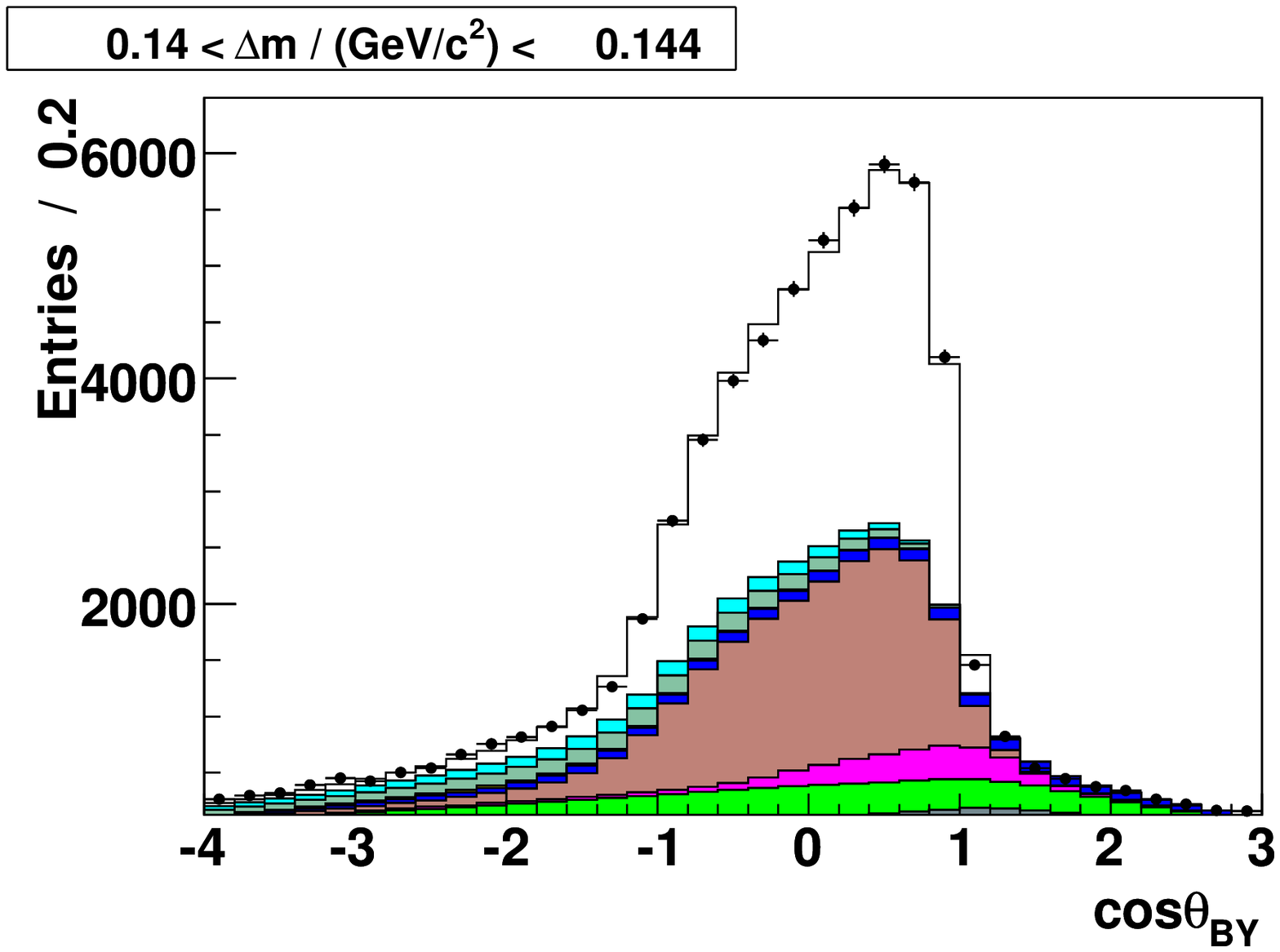}
\put(-11,10){\tiny{$*$}}
\put(-20,145){(b)}
\put(-165,120){\lbabar\ }
\put(-170,110){{preliminary} }
\end{minipage}
\begin{minipage}{0.43\textwidth}
\includegraphics[width=1.0\textwidth,clip,bb=  20 0 514 349]{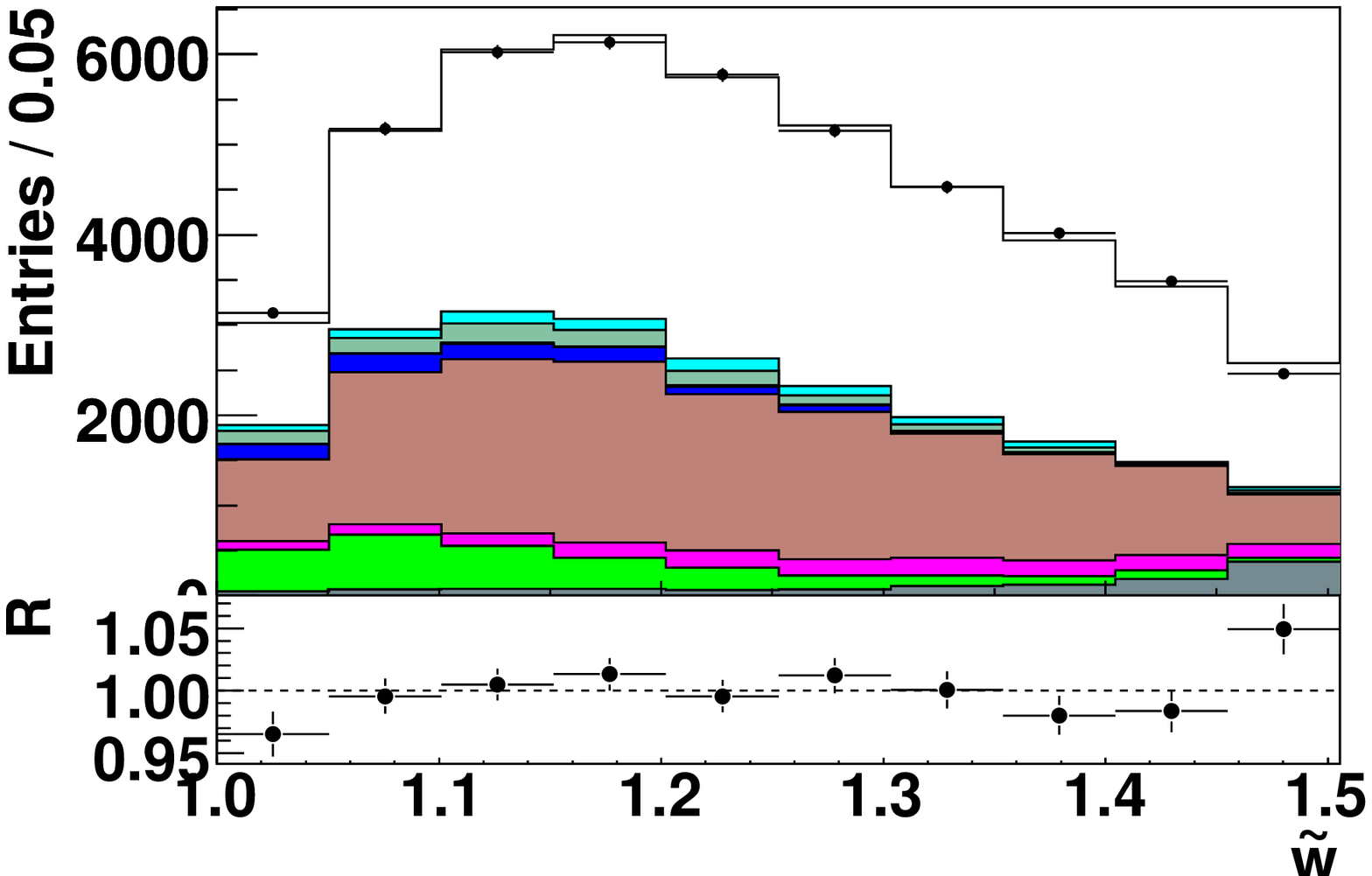}%
\put(-20,135){(c)}
\put(-198,46){\textcolor{white} {$\blacksquare$}}
\put(-145,120){\lbabar\ }
\put(-150,110){{preliminary} }
\end{minipage}
\begin{tabular}{lllcll}
 \vphantom{VVVVVVV}&
 $\square${}                         & Signal                   &\vphantom{VVV}&
 \textcolor{Semisig}{$\blacksquare$} & Signal like              \\
 &\textcolor{Dssdmp} {$\blacksquare$} & $D^{**}$ (\dm-peaking)   & &
 \textcolor{Dlnu}   {$\blacksquare$} & \GCXDzlnu                \\
 &\textcolor{Dssdmf} {$\blacksquare$} & $D^{**}$ (\dm-flat)      & &
 \textcolor{Combds} {$\blacksquare$} & Combinatorial \Dstarz    \\
 &\textcolor{Cor}    {$\blacksquare$} & Correlated               & &
 \textcolor{Ccbar}  {$\blacksquare$} & \ccbar events            \\
 &\textcolor{Uncor}  {$\blacksquare$} & Uncorrelated             & & & 
\end{tabular}
\caption
{
Data distributions (dots with error bars) and fit results
(stacked histograms) for (a) \dm in the \cosby signal range (-1,+1), 
(b) \cosby in the \dm signal range (140,144\mevcc), and
(c) \wtilde in both signal ranges. The plot below (c) shows the ratio fit/data.
The different contributions to the fit function are explained in the
text.
}
\end{center}
\label{fig:dm_cosby_wtilde}
\end{figure}

Figure \ref{fig:dm_cosby_wtilde} shows the result of the fit together with the selected data. The ``Signal'' 
part of the fit function contains the correctly reconstructed 
\BmtoDstarzennueb decays. The two $D^{**}$ parts contain 
$B\to{}D^{**}\electron\nu$ decays with (``\dm peaking'') and without (``\dm flat'') a
correctly reconstructed \Dstarz intermediate state 
($D^{**}=D_{1},D^{*}_{0},{D'}_{1},D^{*}_{2},D^{*}\pi,D\pi$). 
Events with a correctly reconstructed
\Dstarz and a correctly identified electron from the same \B and from
two different \B mesons are in the ``Correlated'' and ``Uncorrelated''
background parts, respectively. ``Signal like'' are true decays
\BmtoDstarzennueb and \BzbtoDstarpennueb which are not correctly
reconstructed. The background from true $\B\to\Dz\electron\nu$ decays
is called ``$\Dz\electron\nu$''.  All other background candidates from
\BB events (``Combinatorial \Dstarz{}'') are flat in the \dm and the
\cosby distribution since they do neither contain a correctly
reconstructed \Dstarz nor do they come from a charmed semileptonic
decay. 
The last contribution, only visible in the highest 
\wtilde bins, comes from \ccbar events in the continuum.

\begin{table}
\begin{center}
\caption
{Summary of input parameters. }
\begin{tabular}{lll}
\hline\hline
{\bf{Input Parameter}}&
{\bf{Value}}&
{\bf{Ref.}}\\
\hline
$\BrFr{\FourS\to\BpBm}$      &  $(50.6 \pm 0.8)\%$     & \cite{pdg:2006} \\     % used in my MC-mix: 50.0  \%
$\BrFr{\Dstarz\to\Dz\piz}$   &  $(61.9 \pm 2.9)\%$     & \cite{pdg:2006} \\     % EvtGen: 61.9  \%
$\BrFr{\Dz\to\Km\pip}$       &  $(3.80 \pm 0.07)\% $   & \cite{pdg:2006} \\     % EvtGen:  3.83 \%
$\BrFr{\piz \to \gaga}$      &  $(98.798 \pm 0.032)\%$ & \cite{pdg:2006} \\     % EvtGen: 98.8  \%
$\tau_{\Bm}$                 &  $(1.638\pm 0.011)\ps$  & \cite{pdg:2006} \\
$R_1(1)$                     &  $1.417 \pm 0.075$      & \cite{Aubert:2006mb}\\
$R_2(1)$                     &  $0.836 \pm 0.043$      & \cite{Aubert:2006mb}\\
\hline\hline
\end{tabular}
\label{tbl:result:input_parameters}
\end{center}
\end{table}

The systematic uncertainties are divided into analysis-internal and
analysis-external ones, see Table~\ref{tbl:systematics:all}. The former are specific to our analysis,
the latter enter by input parameters taken from other
measurements.
Starting with the internal ones,
the relative uncertainty on the efficiency to find a charged
particle's track is 0.8\%, leading to 2.4\% and 1.2\% for ${\cal B}$ and $V$.
The dependence of the tracking efficiency on the transverse momentum
\pt has an uncertainty which could distort the shape of the \wtilde
spectrum.
\begin{table}
\begin{center}
\caption
{Summary of relative systematic uncertainties in percent.}
\begin{tabular}{lccc}
\hline
\hline
\\[-2.4ex]
\ &
$\boldsymbol{{\Delta{}V}/{V}}$ &
$\boldsymbol{{\Delta\rho^2}/{\rho^2}}$ &
$\boldsymbol{{\Delta\cal B}/{\cal B}}$ \\
\\[-2.4ex]
\hline
tracking efficiency ($\epsilon_{\rm tr}$)&
1.2 &
-   &
2.4 \\
\pt dependence of $\epsilon_{\rm tr}$&
0.3 &
0.5 &
0.2 \\
particle ID efficiency&
0.9 &
2.0 &
1.6 \\
extrapolated \piz efficiency ($\epsilon_{\piz}$)&
1.8 &
- &
3.6 \\
$p_{\piz}$ dependence of $\epsilon_{\piz}$&
1.0 &
3.5 &
0.4 \\
\dm shape of $D^{**}${} background &
0.1 &
0.1 &
0.2 \\
shape parameters&
1.0 &
2.5 &
0.6 \\
number of \BB events&
0.6 &
- &
1.1 \\
off-peak luminosity &
0.1 &
0.4 &
$<$0.1 \\
\hline
\bf{total internal} &
$\boldsymbol{2.8}$ &
$\boldsymbol{4.8}$ &
$\boldsymbol{4.8}$ \\
\hline
$R_1(1)${} and $R_2(1)$ &
0.1 &
4.7 &
0.3 \\
$\BrFr{\FourS\to\BpBm}${} &
0.8 &
- &
1.6 \\
$\BrFr{D^{*0} \to \Dz \piz}${} &
2.3 &
- &
4.7 \\
$\BrFr{\Dz \to K^- \pi^+}${} &
0.9  &
- &
1.8 \\
$B^{-}${} life time &
0.3 &
- &
- \\
$D^{**}${} decay fractions &
0.3 &
0.7 &
0.3 \\
number of \Dstarz in \ccbar events&
0.2 &
0.7 &
$<$0.1  \\
\hline
\bf{total external} &
$\boldsymbol{2.6}$ &
$\boldsymbol{4.8}$ &
$\boldsymbol{5.3}$ \\
\hline 
\bf{total } &
$\boldsymbol{3.9}$ &
$\boldsymbol{6.8}$ &
$\boldsymbol{7.2}$ \\
\hline 
\hline 
\end{tabular}
\label{tbl:systematics:all}
\end{center}
\end{table}

The uncertainties arising from the identification of charged tracks
as electrons or as kaons contribute to the result as listed under
``particle ID efficiency''.
A significant fraction of the total uncertainty of our result comes
from the precision of the \piz reconstruction efficiency
($\epsilon_{\piz}$). It is
determined from $\epem\to\tautau$ events where one of the two $\tau$
leptons is either reconstructed
by one track and two clusters (mainly $\tau\to\rho(\pi\piz)\nu$) or it
is reconstructed by only one track without clusters (mainly
$\tau\to\pi\nu,\mu\nu\nub$) \cite{pi0_Effi}. The other $\tau$, used as a $\tau${}-pair tag, 
is reconstructed in the channel $\electron\nu\nub$.
From the numbers of \tautau events reconstructed in each of the two
channels we derive an efficiency in data and in MC,
giving a correction to the simulated \piz efficiency.
The correction is obtained for momenta above 350\mevc and has a
precision of 3\%. 
We add to this value 2\% in quadrature, which is our uncertainty estimate for the extrapolation
to the lower-momentum range with all \piz mesons from $\Dstarz\electron\nu$ decays.
From fit results for different cuts on  $p_{\piz}$ we estimate the
uncertainty in the shape of the \wtilde spectrum which gives one of
the major contributions to the uncertainty of $\rho^2$
(``$p_{\piz}$ dependence of $\epsilon_{\piz}$'').
Corrections to the \dm shape and to the \cosby shape are described by
a parametrization of the \wtilde dependence which also contributes to the
final uncertainties, see ``shape parameters''.
The determination of the total number of \BB events in the
analyzed data sample has a relative uncertainty of 1.1\%. It
contributes only to $V$ and $\cal B$ but not to $\rho^2$.
The uncertainty on the luminosity of the off-peak data sample
propagates also to the final result.

The dominant contribution to the external uncertainty on $\rho^2$
comes from $R_1(1)$ and $R_2(1)$. We determine the derivatives of our
fit result with respect to $R_1$ and $R_2$. We find the values given
in Table \ref{tbl:deriv} leading to the uncertainties listed in Table \ref{tbl:systematics:all}.
\begin{table}
\begin{center}
\caption
{Derivatives of $V$, $\rho^2$, and ${\cal B}$.}
\begin{tabular}{lccc}
\hline\hline
\phantom{VVVVVVV} & 
$V$  &
$\rho^2  $        &
${\cal B}$       \\
\hline
$\partial{}/\partial{}R_1$ &
-0.00038     &
 +0.0303  &
 +0.00382\\
$\partial{}/\partial{}R_2$ &
-0.00158   &
-1.22 &
+0.00551 \\
\hline\hline
\end{tabular}
\label{tbl:deriv}
\end{center}
\end{table}
The input decay fractions only contribute to $V$ and $\cal B$. An
improvement in the precision of \BrFr{\Dstarz\to\Dz\piz} would
significantly improve our results on $V$ and $\cal B$.
The uncertainty on the lifetime $\tau_{\Bm}$ affects only $V$.
Semileptonic $B$ decays into higher excited
charmed mesons, $\B \to D^{**} e \nu$, contribute to the final
uncertainties mainly due to their less precisely known decay fractions
but also due to their description in the fit.
The uncertainty in the number of correctly reconstructed \Dstarz
mesons in $\epem\to\ccbar$ events influences $\cal B$ by less than 0.1\%.

Adding all systematic errors in quadrature leads to the last line in
Table \ref{tbl:systematics:all} and to our preliminary results
\begin{gather*}
\FoFaF(1)\cdot\Vcb=(36.3\pm 0.6\pm 1.4)\cdot 10^{-3}\ ,\\
\rhosqraone= 1.15\pm 0.06\pm 0.08\ , \\
{\cal B}(B^-\to D^{*0} e^-\overline\nu_e) = (5.71\pm 0.08\pm 0.41)\%\ .
\end{gather*}
The correlation coefficients between $\FoFaF(1)\cdot\Vcb$ and 
$\rhosqraone$ are $\varrho=+0.90$ for statistics, +0.43 for systematics,
and +0.52 in total. 

Using $F(1)=0.919\pm 0.033$ from lattice QCD
\cite{Hashimoto:2001nb},
we obtain $\Vcb=(39.5\pm 0.6\pm 2.0)\cdot 10^{-3}$ in good agreement
with the average from the exclusive neutral \B decays $\Bz\to\Dstarm\ellp\nu$,
$(39.2\pm 0.7\pm 1.4)\cdot 10^{-3}$
\cite{Barberio:2007cr},
and in agreement with results from the inclusive decays $\B\to{}X_{c}\ell\nu$,
e.\ g.\ $(42.0\pm 0.2\pm 0.7)\cdot 10^{-3}$ in Ref.\ \cite{Buchmueller}. Our result for
$\rho^2$ is in the center of the range (0.5, 1.5) from the $\Bz\to\Dstarm\ellp\nu$
experiments \cite{Barberio:2007cr}. 
\begin{figure}
\includegraphics[width=0.5\textwidth]{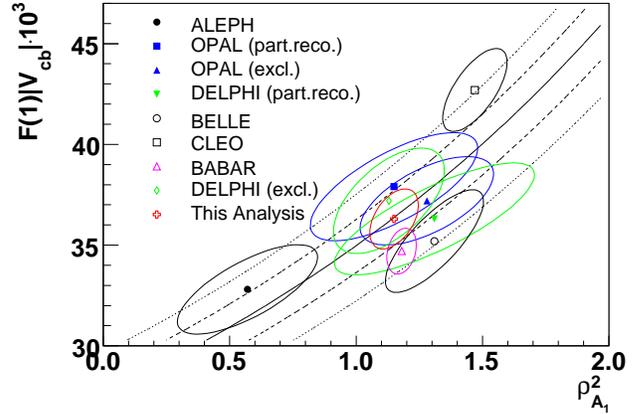}%
\caption{One-standard-deviation contours (stat. and sys. combined) of
$V$ vs. $\rho^2$ \cite{Barberio:2007cr} from all recent \BzbtoDstarpennueb results
 together with our preliminary result. The lines correspond
to constant decay fractions
$\BrFr{\BmtoDstarzennueb}=5.71\%\pm{}1\sigma$ and $\pm{}2\sigma$ with $\sigma=0.42\%$.}
\label{fig:Vcb_vs_rhosqr}
\end{figure}

Figure \ref{fig:Vcb_vs_rhosqr}  shows the $1\sigma$ contour of our result in the 
$\rho^2,\FoFaF(1)\Vcb$ plane together with the contours of the neutral-\B decays.
The quasi-diagonal lines in this Figure are lines of constant
decay fraction $\cal B$.

For a comparison of our decay-fraction result with the decay 
fraction of the neutral-\B decay mode, we use the 
lifetime ratio 
$\tau(\Bp)/\tau(\Bz)=1.076\pm 0.008$ and  
${\cal B}(\Bz\to\Dstarm\ellp\nu)=(5.28\pm{}0.18)\%${} \cite{Barberio:2007cr}. From this, we expect 
${\cal B}(\Bm\to\Dstarz\ellm\overline\nu)=(5.68\pm 0.20)\%$, again in good 
agreement with our result. On the other hand, our decay-fraction result is
about 1.6 $\sigma$ lower than the PDG average 
\cite{pdg:2006}
of the \Bm results from CLEO and ARGUS \cite{Adam:2002uw,Albrecht:1991iz}.

% Input the pubboard acknowledgements file
%\input pubboard/acknow_PRL.tex
We are grateful for the excellent luminosity and machine conditions
provided by our \pep2\ colleagues, 
and for the substantial dedicated effort from
the computing organizations that support \babar.
The collaborating institutions wish to thank 
SLAC for its support and kind hospitality. 
This work is supported by
DOE
and NSF (USA),
NSERC (Canada),
CEA and
CNRS-IN2P3
(France),
BMBF and DFG
(Germany),
INFN (Italy),
FOM (The Netherlands),
NFR (Norway),
MIST (Russia),
MEC (Spain), and
STFC (United Kingdom). 
Individuals have received support from the
Marie Curie EIF (European Union) and
the A.~P.~Sloan Foundation.

% the bibliography


\begin{thebibliography}{99}
\bibitem{Charge_Conj_Decays}
  Charge-conjugated partners are always implied in this text.
\end{thebibliography}
\end{document}